\documentclass[pre,preprint,preprintnumbers,amsmath,amssymb]{revtex4}
\usepackage{graphics,epsfig,epstopdf,stmaryrd}
\newcommand{\mV}{\mathcal{D}}
\newcommand{\mE}{\mathcal{E}}
\begin {document}
\title{Nonlocal adiabatic theory. II. Nonlinear frequency shift of an electron plasma wave in a multidimensional inhomogeneous plasma}
\author{Didier B\'enisti}
\email{didier.benisti@cea.fr} 
\affiliation{ CEA, DAM, DIF F-91297 Arpajon, France.}
\date{\today}
\begin{abstract}
In this article, we provide a general derivation of the nonlinear frequency shift, $\delta \omega$, for a sinusoidal electron plasma wave (EPW) that varies slowly enough for the results derived in the companion paper, on the action distribution function, to apply. We first consider the situation when the EPW monotonously  grows and then monotonously  decays in a homogeneous plasma. In this situation, we show a hysteresis in the wave frequency, which does not converge back to its linear value as the wave decays to small amplitudes. We then address the derivation of $\delta \omega$ for an EPW that keeps growing in a one-dimensional (1-D) inhomogeneous plasma. We show that, usually, the frequency shift does not only depend on the local EPW amplitude and wavenumber. It also depends on the whole history of the density variations, as experienced by the wave. In a multidimensional inhomogeneous plasma, the values assumed by $\delta \omega$ are usually different from those derived in 1-D because, due to the transverse electron motion, one must account for the hysteresis in $\delta \omega$  in addition to plasma inhomogeneity. Hence, unless the EPW keeps growing in a homogeneous one-dimensional plasma, one cannot derive $\delta \omega$ \textit{a priori} as a function of the local wave amplitude and wavenumber. Due to the nonlocality in the action distribution function, $\delta \omega$ depends on the whole history of the variations of the EPW amplitude and plasma density.
\end{abstract}
\maketitle
\section{Introduction}
The derivation of the nonlinear frequency shift, $\delta \omega$, of an electron plasma wave (EPW) has been a long standing issue which  is of fundamental interest, and which also has important implications in several problems relevant to applied plasma physics. In this paper, we generalize previous results on $\delta \omega$ by accounting for plasma inhomogeneity and multi-dimensional effects which, to the best of our knowledge, has never been done before. Although quite general, our theory is designed to hold for physics parameters relevant to the application that motivated the present work, stimulated Raman scattering (SRS) in a fusion plasma. SRS is still a serious concern for inertial fusion, since large and unexpected Raman reflectivities have been measured at the National Ignition Facility~\cite{NIC}, while a robust model able to predict such reflectivities is still missing. Now, the accurate estimate of the nonlinear frequency shift of an SRS-driven plasma wave is of prime importance to model stimulated Raman scattering in the nonlinear kinetic regime. Indeed, depending on its variations, $\delta \omega$ may either induce a phase mismatch that leads to SRS-saturation~\cite{casa,friou}, or may compensate the detuning resulting from plasma inhomogeneity~\cite{benisti10b,chapman}, and let SRS grow more efficiently that linear theory~\cite{rosenbluth}~would predict.  Moreover, due to its transverse profile, the nonlinear frequency shift entails the bending of the EPW phase front which, in turn, leads to the self-focusing of the wave and saturates SRS~\cite{yin,SRS3D}.  

For the typical parameters of laser-plasma interaction in a fusion device, the EPW varies so slowly that one may use the practical formulas obtained in the companion paper~\cite{fadia} to derive the electron distribution function. Moreover, SRS usually grows and saturates so quickly that the ion motion may be neglected~\cite{rousseaux}. The slow variations in the distribution function of the passing particles, as given by Eq.~(98) of the companion paper~\cite{fadia},~are also usually negligible, so that one may just rely on adiabatic formulas.  Indeed, in a homogeneous one-dimensional (1-D) plasma, they proved to yield values for~$ \delta \omega$ in very good agreement with those inferred from Vlasov simulations of SRS~\cite{benisti08}. In particular, we insist here on the fact that, in the numerical simulations of Ref.~\cite{benisti08}, the EPW amplitude and phase velocity were space-dependent, yet, adiabatic results were very accurate. Furthermore, relativistic corrections to the electron motion proved to be negligible, and the EPW may usually be considered sinusoidal. Hence, the adiabatic results derived in our companion paper as regards the time variations of the action distribution, namely Eqs.~(88)-(94) of Ref.~\cite{fadia}, should directly apply, and they will be indeed used to derive $\delta \omega$.

Although we restrict to a slow evolution of the EPW, the present work generalizes many previous ones, which we now discuss. One of the best known paper on the nonlinear frequency shift is that published by Morales and O'Neil~\cite{morales}, where the authors assumed that a sinusoidal EPW had suddenly grown from zero to a finite constant value, in a 1-D homogeneous plasma. They showed that, after a few oscillations, the EPW frequency eventually reached a constant value, which they calculated analytically by assuming that the EPW amplitude and phase velocity had remained constant (thus neglecting the first oscillations in the EPW frequency). Karpman, Istomin and Shklyar (KIS) generalized the calculation of Morales and O'Neil in Refs.~\cite{kis1,kis2}  by deriving the phase-mixed value of $\delta \omega$ (i.e., that obtained after the oscillations have damped away) 
in a non-uniform plasma. Then, in addition to the result derived by Morales and O'Neil, KIS found an extra term in $\delta \omega$ due to plasma inhomogeneity, for which they provided an analytic expression. Clearly, the derivation of $\delta \omega$ by Morales and O'Neil and by KIS differ from ours because they made use of the sudden approximation instead of the adiabatic one. While we calculate the population of the trapped and untrapped electrons  for a wave that varies slowly in space and time, the aforementioned authors derived the nonlinear electron distribution function by assuming that the wave amplitude remained constant. Clearly, by making use of such a hypothesis they could not address the nonlocality in $\delta \omega$, which is the main point of the present paper.

In Ref.~\cite{dewar}, Dewar derived $\delta \omega$ in the situation when the time evolution of the EPW was slow enough for the electron motion to be adiabatic. Moreover, he assumed that the EPW kept growing in 1-D uniform, and initially Maxwellian, plasma. Then, he provided an estimate for $\delta \omega$ proportional to the square root of the EPW amplitude (in the limit of small amplitudes), which he also derived by neglecting the continuous evolution of $\delta \omega$ as the wave grew. Recently, Dewar's calculation has been improved by Liu and Dodin~\cite{liu}, who used exactly the same hypotheses but expressed $\delta \omega$ as the sum of term proportional to the square root of the amplitude and a term proportional the the amplitude squared. However, as discussed in the companion paper~\cite{fadia}, the adiabatic distribution function at a given time depends on all previous values of the wave phase velocity. Hence, the adiabatic derivation of the frequency shift should account for the continuous change in the EPW frequency. In this paper, when deriving $\delta \omega$ we do account for the continuous change in the wave phase velocity (due to nonlinearity and plasma inhomogeneity), and for the nonlocality in the electron distribution function it entails, which is one major difference with Refs.~\cite{dewar,liu}.

Moreover, an EPW may only grow in an initially Maxwellian plasma if it has been externally driven, and one needs to account for the external drive in the dispersion relation. In the nonlinear regime, this is needed even when the wave has been driven by an electrostatic potential locally, and then freely propagates. Indeed, the external potential has contributed to the building of the population of trapped and untrapped electrons. Consequently,  it cannot be ignored when deriving the boundary condition corresponding to the nonlinear electron distribution, and the associated EPW, at the location when one may assume that the external potential has become negligible. For a laser-driven wave, the situation is even more complicated because the external drive is to be accounted for at all space locations. Consequently, unlike what has been done in most papers on the subject, e.g. Refs.~\cite{morales,kis1,kis2,dewar,liu}, one may usually not resort to the free dispersion relation, $1+\chi_r=0$, where $\chi_r$ is the real part of the electron susceptibility. For an SRS-driven wave, the dispersion relation reads $1+\alpha_d \chi_r=0$, where $\alpha_d$ (whose definition may be found in Ref.~\cite{benisti08}) is larger than unity, which may be understood as follows. In the linear regime, the SRS-growth rate $\gamma_{SRS}$ is proportional to $[(1+\chi_r)^2+(\nu_L \partial_\omega \chi_r)^2]^{-1}$, where $\nu_L$ is the Landau damping rate of the driven EPW. Because $\nu_L$ decreases with the wave phase velocity, $\gamma_{SRS}$ reaches its maximum for a frequency, $\omega$, and a wave number, $k$, such that $\omega/k$ is larger than for the natural plasma mode (solving $1+\chi_r=0$). Hence, $\omega$ and $k$ solve $1+\alpha_d \chi_r=0$, with $\alpha_d>1$.   Now, after gain narrowing,  the nearly monochromatic wave that results from SRS is the one that has the largest linear growth rate, and its linear dispersion relation is therefore $1+\alpha_d \chi_r=0$. Clearly, $\alpha_d$ should be very close to unity when $\nu_L$ is small, which is true for small enough values of $k \lambda_D$, $\lambda_D$ being the Debye length. As shown in Ref.~\cite{benisti08} (for a uniform plasma), one may obtain a very accurate estimate of the nonlinear frequency shift of an SRS-drive plasma wave by simply solving $1+\chi_r=0$, whenever $k\lambda_D <0.35$.  At this stage, one may note that the effect of $\alpha_d$ on the EPW frequency is similar to that recently discussed in Ref.~\cite{kaganovich} for the frequency of the SRS-scattered electromagnetic wave. One may also note that, in the nonlinear regime, $\alpha_d$ quickly converges towards unity as the wave grows. However, the latter point, which is detailed in Ref.~\cite{benisti08}, is outside the scope of this paper. Indeed, here we want to focus on the nonlocality in $\delta \omega$ by avoiding any effect (like that due to the drive) which would complicate the dispersion relation. Moreover, we only consider waves that vary slowly enough for adiabatic or neo-adiabatic theory to apply. Consequently, we restrict to situations when the frequency shift may be accurately derived by simply solving $1+\chi_a=0$ where $\chi_a$, which is defined by Eqs.~(\ref{chia})-(\ref{chit}),  is the adiabatic estimate of the electron susceptibility. As discussed above, for an SRS-driven plasma wave this is true provided that the linear value of the wave number is such that $k\lambda_D<0.35$. Moreover, for the sake of simplicity we restrict to sinusoidal EPW's (see Ref.~\cite{lindberg} for a treatment that does not make use of this hypothesis) which  is usually a good approximation for SRS~\cite{benisti08}.

We want to insist here on the fact that we choose an initially Maxwellian plasma, and implicitly consider driven waves, only for definiteness, and because this has been the most widespread choice in previous papers on the subject. However, the equations we derive, their numerical resolution and, most importantly, the main concept  we discuss in this paper i.e., the nonlocality in $\delta \omega$,  do apply to any wave, that would grow or decay, provided that the wave properties vary slowly enough, according to Eq.~(10) or Eq.~(11) of the companion paper~\cite{fadia}. Moreover, as discussed above, we restrict here to wave numbers such that the dispersion relation is not affected by the drive, so that our results may be straightforwardly extended to waves resulting from an electrostatic instability. 

We start by deriving $\delta \omega$ when the wave first monotonously grows in a uniform plasma to an amplitude which is large enough to induce a significant frequency shift, and then monotonously decays back to small amplitudes. In this situation, we show that there is a hysteresis in the wave frequency, which does not converge back to its linear value when the wave amplitude decays to zero, as illustrated in Fig.~\ref{fhys}. This is a direct consequence of the nonlocality in the electron distribution function.  

Then, we compute the nonlinear frequency of an EPW that keeps growing in a one-dimensional (1-D) inhomogeneous plasma. When the plasma is stationary, the wave frequency remains constant in the linear regime, so that the wavenumber has to vary with the density for the dispersion relation to remain fulfilled. Hence, in the linear regime the phase velocity, $v_\phi$, varies because of plasma inhomogeneity. In the nonlinear regime, one also has to account for the nonlinear frequency shift, so that $v_\phi$ varies with the density and with the EPW amplitude. Now, because the action distribution function, $f(I)$, is not local in $v_\phi$~\cite{fadia}, it cannot be expressed as a  function of the local EPW amplitude and plasma density. One of the main points of this paper is to estimate the impact on $\delta \omega$ of this nonlocality in $f(I)$. In particular, we discuss in which situations may $\delta \omega$ be derived in advance, as a function of the local EPW amplitude and wavenumber, regardless of how the wave has grown. This is an important issue to build a code that that would compute  SRS reflectivity with a good efficiency. As illustrated in Figs.~\ref{f6} and \ref{f7},   depending on the physical situation, a local derivation of $\delta \omega$ may be rather accurate, or significantly wrong. 

Finally, we derive $\delta \omega$ in a two-dimensional (2-D) inhomogeneous plasma. In a multi-dimensional geometry, the EPW is located within a given domain, $\mV$, which is assumed to be much more elongated along the EPW direction of propagation than across it. Then, due to their transverse motion, the electrons may quickly cross $\mV$. They first experience an increasing wave amplitude as they enter the domain $\mV$ while, when they exit this domain, they experience a decreasing amplitude. This entails variations in $\delta \omega$ similar to the hysteresis mentioned above, that one needs to account for in addition to plasma inhomogeneity.  Then, depending on how fast the electrons cross $\mV$, 1-D results may be relevant or completely inaccurate,  as illustrated in Figs.~\ref{f8}~and~\ref{f9}.

This article is organized as follows. In Section~\ref{dispersion}, we quickly recall the adiabatic dispersion relation for a sinusoidal electrostatic wave. We first solve this dispersion relation in Section~\ref{hysteresis} for an EPW in a homogeneous plasma whose amplitude monotonously grows and, then, monotonously decays. This allows us to show a hysteresis in the wave frequency, resulting from the nonlocality of the adiabatic distribution function. In Section~\ref{1D}, we derive the nonlinear frequency shift of an EPW that keeps growing in a 1-D inhomogeneous plasma. Moreover, we compare the values of $\delta \omega$ following from the direct resolution of the adiabatic dispersion relation with those obtained by using local formulas. The derivation of the nonlinear wave frequency is generalized in Section~\ref{2D}, for an EPW propagating in a two-dimensional inhomogeneous plasma. In this situation, we discuss the impact on $\delta \omega$ of the hysteresis and of the nonlocality entailed by plasma inhomogeneity. Finally, Section~\ref{conclusion} concludes this work. 

\section{The adiabatic dispersion relation of a sinusoidal electron plasma wave}
\label{dispersion}
In this Section, we quickly recall the results derived in Ref.~\cite{benisti07}~as regards the adiabatic dispersion relation of a sinusoidal EPW. Just like in Ref.~\cite{fadia}, we assume that the EPW electric field reads, $\mE(x,t)=E_0 \sin[\varphi(x,t)]$. Then, the electron dynamics derives from the Hamiltonian,
\begin{equation}
\label{H}
H=Ê(v-v_\phi)^2/2-\Phi \cos(\varphi),
\end{equation}
for the canonically conjugated variables $\varphi$ and $v$, whose evolution is given as a function of the normalized time, $\tau \equiv kv_{th} t$, $v_{th}$ being the longitudinal thermal velocity.  In Eq.~(\ref{H}),  $\Phi \equiv eE_0/kT_e$, $T_e$ being the longitudinal electron temperature, and $k\equiv \partial_x \varphi$ the wavenumber. Moreover, in Eq.~(\ref{H}),  $v_\phi \equiv \omega/kv_{th}$ where the wave frequency, $\omega$, is $\omega \equiv -\partial_t \varphi$. 

Now, directly from Gauss law one easily finds that, about a given normalized position $\varphi_0$,  the dispersion relation of a freely propagating sinusoidal EPW is (see Refs.~\cite{benisti07,benisti16} for details), 
\begin{equation}
\label{2}
1+\frac{1}{\pi(k\lambda_D)^2\Phi} \int \int_{0}^{2\pi}f(I,\varphi_0) \cos[\varphi(\theta,I)] d\theta dI=0,
\end{equation}
where $I$ and $\theta$ and are, respectively, the action and angle variables, and where $f(I,\varphi_0)\equiv \langle \tilde{f}(\theta,I) \rangle$, $\tilde{f}$ being the electron distribution function in action-angle variables normalized to unity, and $\langle \tilde{f} \rangle$ its space average over one wavelength about $\varphi=\varphi_0$.  Actually, as discussed in Ref.~\cite{fadia}, one may only define unambiguously $\tilde{f}(\theta,I)$ within each sub-region of phase space i.e., either above the upper branch of the separatrix [region $(\alpha)$], or below the lower branch [region $(\beta)$], or inside the separatrix [region $(\gamma)$]. Therefore, the $I$-integral in Eq.~(\ref{2}) must be understood as a sum of $I$-integrals over each sub-region.

Using the expression for $\int \cos[\varphi(\theta,I)] d\theta$ derived in Ref.~\cite{benisti07}, the dispersion relation reads, 
\begin{equation}
\label{3}
1+\chi_a=0,
\end{equation}
with,
\begin{equation}
\label{chia}
\chi_a \equiv \chi_u+\chi_t,
\end{equation}
and
\begin{eqnarray}
\label{chiu}
\chi_u & \equiv & \frac{2}{(k\lambda_D)^2\Phi} \int_{v_{tr}+v_\phi}^{+\infty} \left[f_\alpha(I)+f_\beta(I-2v_\phi)\right]\left\{1+\frac{2}{m}\left[\frac{E(m)}{K(m)}-1\right]\right\}dI \\
\label{chit}
\chi_t & \equiv &\frac{2}{(k\lambda_D)^2\Phi}  \int_0^{v_{tr}} f_\gamma(I) \left\{-1+2\frac{E(m)}{K(m)}Ê\right\} dI.
\end{eqnarray}
In Eq.~(\ref{chiu}), $m$ is related to $I$ by,
\begin{equation}
\label{5}
I=v_{tr}E(m)/\sqrt{m},
\end{equation}
where $v_{tr}Ê\equiv 4\sqrt{\Phi}/\pi$, and where $E(m)$ is the elliptic integral of second kind~\cite{abramowitz}. In Eq.~(\ref{chit}),
\begin{equation}
\label{6}
I=v_{tr}Ê\left[E(m)+(m-1)K(m)Ê\right],
\end{equation}
where $K(m)$ is the elliptic integral of first kind~\cite{abramowitz}. As for $f_\alpha$(I), $f_\beta(I)$ and $f_\gamma(I)$ they are calculated by using the procedure given in the companion paper i.e., by using Eqs.~(88)-(94) of Ref.~\cite{fadia}. Then, $\chi_a$ is perfectly defined, and one just has to solve Eq.~(\ref{3}) to derive the EPW nonlinear frequency shift. However, there are several caveats in the actual computation of $\chi_a$, which are detailed in the following Sections. Before entering into these details, we want to underline the maint point of the paper, i.e., the fact that the dispersion relation (\ref{3}) is not local because, as shown in the companion paper~\cite{fadia}, the distribution functions $f_\alpha(I)$, $f_\beta(I)$ and $f_\gamma(I)$ are not local. They depend on all previous values of the wave phase velocities, and on all the previous history of separatrix crossings by electrons orbits, i.e., on the whole history of electron trapping or detrapping. In particular, the distribution functions depend on all previous minima and maxima of the wave amplitude. $f_\alpha$(I), $f_\beta(I)$ and $f_\gamma(I)$~will not be the same,  and therefore $\chi_a$ will not be the same, if the amplitude keeps growing to  $\Phi_0$, or if it grows to $\Phi_1>\Phi_0$ and then decays to $\Phi_0$.  Indeed, in the first case the electrons have kept on being trapped while, in the second case, the electrons have been trapped and detrapped. Now, as shown in the companion paper~\cite{fadia}, and as is clear from Eqs.~(\ref{7}) and (\ref{7b}) of Section~\ref{hysteresis}, detrapping changes the distribution of untrapped electrons compared to its initial value. This is at the origin of the hysteresis which we are now going to discuss. 

\section{Hysteresis of the wave frequency in a one-dimensional uniform plasma}
\label{hysteresis}

\subsection{Hypotheses}
In this Section, we solve Eq.~(\ref{3}) for an EPW that first monotonously grows from a nearly null amplitude, and then monotonously decays back to small amplitudes. Moreover, we assume that the wavenumber remains constant. Actually, even in a homogeneous plasma, the wavenumber is expected to vary with time. Indeed, due to the nonlinear frequency shift, the wave frequency depends on the wave amplitude which is usually space-dependent. Hence, the wave frequency is not homogeneous and, due to the consistency relation $\partial_tk=-\partial_x \omega$, the wavenumber varies with time. However, in 1-D, the space variations of $\omega$ are usually small enough to neglect the changes they induce on $k$, at least within the time it takes to an SRS-driven plasma wave to grow up to saturated levels, for parameters relevant laser fusion.  This has been shown in Ref.~\cite{benisti08} where the adiabatic values of $\delta \omega$ derived by assuming a constant wavenumber were in excellent agreement with those inferred from 1-D Vlasov simulations of SRS. 

Moreover, we assume that the EPW grows in an initially Maxwellian plasma. This is only possible if the wave is driven, while the dispersion relation Eq.~(\ref{3}) is, \textit{a priori},~only valid for a freely propagating wave. However, as shown in Ref.~\cite{benisti08} and discussed again in the Introduction, the effect of the drive on the wave frequency is negligible when $k\lambda_D \alt 0.35$ so that, for the sake of simplicity, we restrict to such small wavenumbers. 
\subsection{Hysteresis in the wave frequency}
\begin{figure}[!h]
\centerline{\includegraphics[width=12cm]{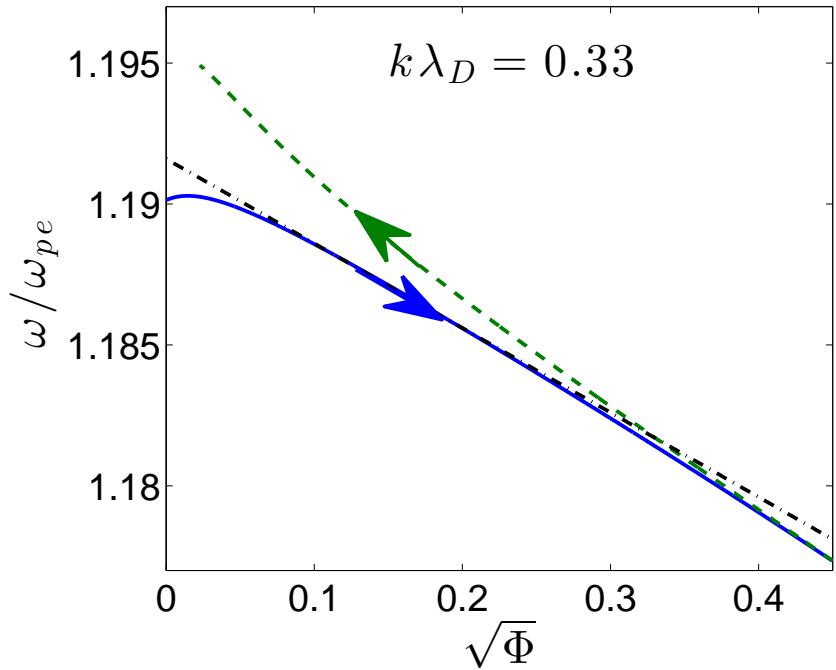}}
\caption{\label{fhys} (Color online) Amplitude dependence of the EPW  frequency when $k\lambda_D=0.33$, and when the amplitude increases (blue solid line), or when it decreases (green dashed line). The arrows indicate the time variation of the wave amplitude. The black dashed-dotted line plots $\delta \omega/\omega_{pe}$ as given by Eqs.~(\ref{13}) and (\ref{12}).}
\end{figure}

Fig.~\ref{fhys} illustrates an example of the amplitude variations of the wave frequency, when $k\lambda_D=0.33$. This Figure shows that $\omega$ does not assume the same values as a function of $\Phi$ when the wave amplitude increases as when it decreases. Hence, there is a hysteresis in the wave frequency, which we now explain.

From the results illustrated in Fig.~\ref{fhys}, $dv_\phi/dv_{tr} \ll 1$. Then, as shown in  the companion paper Ref.~\cite{fadia}, the particles are detrapped nearly symmetrically with respect to the phase velocity. More precisely, let us consider electrons with initial action $I$ in region $(\alpha)$, which are trapped at time $t_1$ while the wave is growing. Let us moreover denote by $v^*_\phi(I)$ the value assumed by the wave phase velocity when trapping occurs, and by $v_{tr}^*$ the corresponding value of $4\sqrt{\Phi}/\pi$. Then, electrons initially in region $(\beta)$  with action $I-2v^*_\phi(I)$ are also trapped at time $t_1$ when $v_{tr}=v_{tr}^*$. Moreover, after trapping they lie on the same orbit as the electrons initially in region $(\alpha)$ with action $I$.  When the wave is decaying, the electrons are detrapped at the time $t_2$ when $v_{tr}$ has decreased back to $v_{tr}^*$, and we denote by $v_\phi^{*'}(I)$ the corresponding value of the phase velocity. Then, if we denote by  $f_{\alpha}^<(I)$ [respectively $f_{\beta}^<(I)$] the action distribution function in region $(\alpha$) ([respectively in region $(\beta)$] before trapping, and by $f_{\alpha}^>(I)$ and $f_{\beta}^>(I)$ these distribution functions after detrapping (i.e. when $t> t_2$), we know from the results of Ref.~\cite{fadia}~that,
\begin{eqnarray}
\label{7}
f^{>}_{\alpha}(I) & \approx & \frac{f^<_\alpha[I+v^*_{\phi}(I)-v^{*'}_{\phi}(I)]+f^<_\beta[I-v^*_{\phi}(I)-v^{*'}_{\phi}(I)]}{2}, \\
\label{7b}
f^{>}_{\beta}[I-2v^*_\phi(I)] & \approx & \frac{f^<_\alpha[I+v^{*'}_{\phi}(I)-v^{*}_{\phi}(I)]+f^<_\beta[I+v^{*'}_{\phi}(I)-3v^{*}_{\phi}(I)]}{2}.
\end{eqnarray}
Usually, $f^<_\beta[I-2v^*_\phi(I)] \gg f^<_\alpha(I)$ so that both distribution functions, $f_\alpha$ and $f_\beta$, significantly change after detrapping compared to their initial values.  Note that, since the action remains conserved before trapping, $f_\alpha^{<}(I)$ and $f_\beta^{<}(I)$ are just the initial action distribution function, $f_0(I)$. Note also that, from Eq.~(\ref{chiu}), $\chi_u$ remains constant provided that $f_\alpha(I)+f_\beta(I-2v_\phi)$ does not change, where $v_\phi$ is the current phase velocity. Using Eqs.~(\ref{7}) and (\ref{7b}), we find that, 
\begin{eqnarray}
\nonumber
f^{>}_{\alpha}(I)+f^{>}_{\beta}(I-2v_\phi) &=& \frac{f_0[I+v^*_{\phi}(I)-v^{*'}_{\phi}(I)]+f_0[I+v^{*'}_{\phi}(I)+v^{*}_{\phi}(I)-2v_\phi]}{2} \\
\label{7c}
&&+\frac{f_0[I-v^*_{\phi}(I)-v^{*'}_{\phi}(I)]+f_0[I+v^{*'}_{\phi}(I)-v^{*}_{\phi}(I)-2v_\phi]}{2}.
\end{eqnarray}
Since $v_\phi^*(I)\approx v_\phi^{*'}(I) \approx v_\phi$ we conclude that, even though $f_\alpha(I)$ and $f_\beta(I)$ might change a lot due to detrapping, $\chi_u$ does not change much. The impact of detrapping on the wave frequency remains modest.  However, because $f_\alpha(I)+f_\beta(I-2v_\phi)$ is not exactly conserved, from Eq.~(\ref{chiu}) $\chi_u$ and therefore $\chi_a$ are not conserved either. For the same amplitude, $\Phi_0$, the EPW dispersion relation is not the same when the wave amplitude has kept growing to $\Phi_0$ as when it has grown to $\Phi_1>\Phi_0$ and then decayed to $\Phi_0$. Consequently, there is a hysteresis in the EPW frequency, $\omega$, which is not negligible compared to $\delta \omega$. Indeed, for the situation considered in Fig.~\ref{fhys}, when $\Phi$ decreases back to small values, $\omega$ differs from its linear limit by more than one third of the maximum frequency shift. 

A hysteresis in the wave frequency has  been observed numerically in Ref.~\cite{berger} for $k \lambda_D=1/3$ using Vlasov simulations. In these simulations,  an EPW grew in an initially Maxwellian plasma under the action of an externally imposed drive. Moreover, periodic boundary conditions were chosen, so that $k$ remained constant. However, the effect of the hysteresis was opposite to the one plotted in Fig.~\ref{fhys}. For the one simulation result presented in Ref.~\cite{berger}, when the EPW first monotonously grew and then decayed back to $\sqrt{\Phi}\approx 0.3$, the wave frequency was found smaller than when the EPW had kept on growing. Now, as indicated in the text, the drive history was not ``ideally adiabatic in the simulations''. This might explain the discrepancy with our theoretical results. 

\subsection{Numerical check of the adiabatic estimates}
We checked numerically our estimate for the nonlinear variations of the wave frequency by making use of test particles simulations. From Eq.~(\ref{2}), we know that when the EPW dispersion relation is satisfied, the statistical averaged value of $\cos(\varphi)$ is $-(k\lambda_D)^2/2$. Consequently, if we let $N$ electrons be acted upon by a slowly varying electrostatic wave whose frequency solves Eq.~(\ref{2}), the averaged value of $\cos(\varphi)$ over all electrons should be $-(k\lambda_D)^2/2$. Then, our numerical check for the nonlinear variations of the EPW frequency plotted in Fig.~\ref{fhys} is very simple.  Numerically, we solve the equations of motion derived from the Hamiltonian $H$ defined by Eq.~(\ref{H}), for $N$ initial conditions consistent with a Maxwellian distribution function. Moreover, we let the wave amplitude vary slowly enough for the electron motion to be nearly adiabatic, and we let $v_\phi$ change with the amplitude in a way that is consistent with the nonlinear variations of $\omega$ plotted in Fig.~\ref{fhys}. Then, we compute,
\begin{equation}
\label{10}
k_{num}Ê\lambda_D\equiv \sqrt{-2\sum_{i=1}^N p_i\cos(\varphi_i)},
\end{equation}
where $\varphi_i$ is the normalized position of the $i^{th}$ electron and $p_i$ is weight. In practice, we use $p_i=e^{-v_{0_i}^2/2}/\sum_{j=1}^N e^{-v_{0_j}^2/2}$, $v_{0_i}$ being the initial velocity of the $i^{th}$ electron. If the frequencies plotted in Fig.~\ref{fhys} indeed solve the dispersion relation Eq.~(\ref{2}), the  right-hand side of Eq.~(\ref{10}) should be a constant, $k_{num}Ê\lambda_D=0.33$.  

\begin{figure}[!h]
\centerline{\includegraphics[width=14cm]{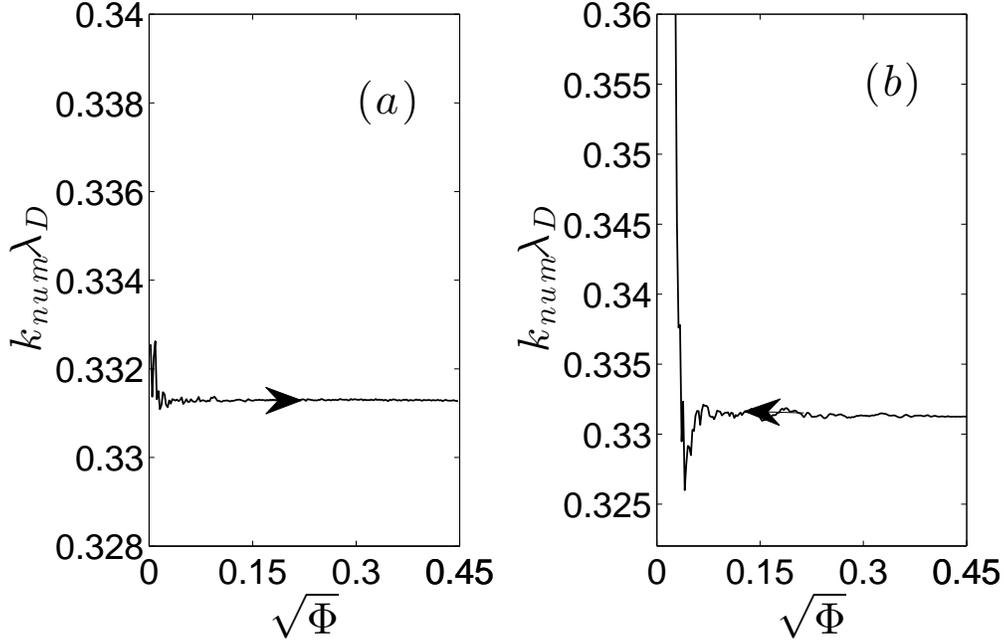}}
\caption{\label{knum} Values of $k_{num}\lambda_D$ obtained for the values of $\omega/\omega_{pe}$ plotted in Fig.~\ref{fhys}, panel (a), when the wave is growing and, panel (b), when the wave is decaying (the arrows indicate how the wave amplitude varies with time).}
\end{figure}

Numerically, our initial velocity distribution function is, 
\begin{equation}
\label{11}
f_0(v)=\frac{1}{\sqrt{2\pi}} \sum_{i=-N_v}^{N_v} e^{-v_i^2/2}\delta(v-v_i),
\end{equation}
where $v_i \equiv v_{\phi_{lin}}+i\delta v$, $v_{\phi_{lin}}\approx 3.6$ being the linear value of the phase velocity, $\delta v=10^{-4}$, and $N_v=6.4\times10^{4}$. Moreover, for each value of $v_i$ we choose 16 initial positions, evenly spaced between $-\pi$ and $\pi$. 

As regards the time evolution for $\Phi$, we first let it grow exponentially in time, 
\begin{equation}
\label{11}
\Phi=\Phi_0e^{\gamma \tau}
\end{equation}
with $\Phi_0=10^{-10}$ and $\gamma=10^{-3}$. Then, after the time $\tau_{\max}$ when $\Phi$ has reached the value $\Phi_{\max}=0.45^2$, we let $\Phi$ decrease exponentially in time, $\Phi=\Phi_{\max}e^{\gamma(\tau_{\max}-\tau)}$. 

Fig.~\ref{knum} unambiguously shows that $k_{num}\lambda_D$ always remains close to $0.33$, whether the wave is growing or decaying. This confirms our theoretical calculations and, in particular, the hysteresis in the wave frequency.  Note, however, that when the wave is growing, $k_{num}\lambda_D \approx0.3313$ instead of exactly $0.33$, because the adiabatic approximation is just an approximation. When the wave amplitude is decreasing, the values of $k_{num}\lambda_D$ are more noisy than when it is increasing. This is because our initial distribution function, $f_0(v)$ defined by Eq.~(\ref{11}), is not smooth. Consequently, the detrapping probabilities do not exactly follow the theoretical ones, as discussed in Ref~\cite{fadia}. However, our choice for $f_0$ is vindicated but the fact that it requires fewer initial positions, than with a smooth distribution function, to yield accurate results. 

More interestingly, we also find that $k_{num}Ê\lambda_D$ seems to diverge from the expected value, $k_{num}Ê\lambda_D=0.33 $, when $\Phi \rightarrow 0$. Actually, when the wave amplitude decreases, the adiabatic dispersion relation cannot be solved down to $\Phi=0$, and we now explain why.

\subsection{Limitations of the adiabatic dispersion relation}
There are caveats in the derivation the adiabatic susceptibility, $\chi_a$, in the limit of small amplitudes, which are detailed in the Appendix. In this Paragraph, we only summarize the corresponding results. 

\subsubsection{Increasing wave amplitude}

As is obvious from Eq.~(\ref{chia}), $\chi_a$ is the sum of the contributions from the trapped electrons and from the untrapped ones. When the wave amplitude increases from $\Phi \approx 0$ and $v_\phi = v_{\phi_0}$, $\chi_t\sim 16f_0(v_{\phi_0})/3\pi\sqrt{\Phi}$, while $\chi_u \sim-16f_0(v_{\phi})/3\pi\sqrt{\Phi}$. Now, as shown in the Appendix, for small wave amplitudes, $v_\phi-v_{\phi_0} \sim (4\eta_v/\pi)\sqrt{\Phi}$, where $\eta_v$ is a constant. Hence, $\chi_a$ converges towards a finite value, $\chi_0$, when $\Phi\rightarrow 0$. 

The law, $v_\phi-v_{\phi_0} \sim (4\eta_v/\pi)\sqrt{\Phi}$, entails,
\begin{equation}
\label{13}
\delta \omega/\omega_{pe}Ê\sim \eta \sqrt{\Phi},
\end{equation}
where, from the results of the Appendix, we know that when $f_0(v)$ is a Maxwellian $\eta$ solves the following equation, 
\begin{equation}
\label{12}
\eta \approx  -\frac{(\omega_{lin}/\omega_{pe})Ê(1.09 +3 \eta^2)f^{''}_0(v_{\phi_0}) }{(\omega_{lin}/\omega_{pe})^2-1-(k\lambda_D)^2-1.2 \eta v_{\phi_{lin}} f^{''}_0(v_{\phi_0})},
\end{equation}
where $\omega_{lin}$ is the linear value of the EPW frequency. Eqs.~(\ref{13}) and (\ref{12}) provide an expression for the frequency shift that accounts for the continuous change in the EPW phase velocity with the amplitude. This improves the  result previously published in Ref.~\cite{benisti08}, where $\delta \omega/\sqrt{\Phi}Ê\omega_{pe}$ was given by the right-hand side of Eq.~(\ref{12}) with $\eta=0$. However, when $k\lambda_D<0.35$, the difference between Eq.~(\ref{12}) and the formula given in Ref.~\cite{benisti08}~is small. 

Fig.~\ref{fhys} shows that Eqs.~(\ref{13}) and (\ref{12}) provide a good approximation of the EPW frequency, at least when $\sqrt{\Phi}<0.45$ and when the wave grows. However, surprisingly enough, for very small amplitudes the agreement with the numerically derived values of $\omega/\omega_{pe}$ is not good. Let us now explain why. As discussed in the Appendix, when $\Phi \rightarrow 0$, $\chi_a$ converges towards a finite limit, $\chi_0$, which is not the linear electron susceptibility $\chi_{lin}$ defined by Eq.~(\ref{A9}), although it is very close to it. In order to derive $\chi_0$, one has to know in advance how $v_\phi$ varies with the wave amplitude in the limit when $\Phi \rightarrow 0$, and the difference between $\chi_0$ and $\chi_{lin}$ is actually proportional to $\eta$. However, when we solve numerically $1+\chi_a=0$, we make no assumption, \textit{a priori},~as regards the amplitude dependence of $v_\phi$. Consequently, when $\Phi=0$, we just solve $1+\chi_{lin}=0$, and we find $v_\phi=v_{\phi_{lin}}$. Then, we let the numerical solution converge towards the adiabatic one. Therefore, numerically, we do not exactly solve $1+\chi_a=0$ for very small wave amplitudes. This explains the difference between the blue solid line and the black dashed-dotted line in Fig.~\ref{fhys} when $\sqrt{\Phi}<0.1$. The black dashed-dotted line is the correct solution to the adiabatic dispersion relation (at least for small amplitudes). However, the blue solid line provides values for $\omega$ which are closer to what is expected in reality. Indeed, it is known that adiabatic results are not valid when $\Phi \approx 0$, and only become accurate once the bounce frequency is of the order of the EPW growth rate. Therefore, the value of the wave phase velocity when $\Phi=0$ is $v_{\phi_{lin}}$. Then, when the wave amplitude increases, $v_\phi$ must smoothly change from $v_{\phi_{lin}}$ to the solution of $1+\chi_a=0$, which is exactly what the blue solid line in Fig.~\ref{fhys} does. Moreover, we checked that we obtained exactly the same curve with 200 values for $\sqrt{\Phi}$ between 0 and 0.45, or with $2\times10^4$ values. Therefore, we believe that the blue line in Fig.~\ref{fhys} describes the transition from the linear EPW frequency to the adiabatic nonlinear one. Physically, it is more relevant than the solution of the adiabatic dispersion relation. 

\subsubsection{Decreasing wave amplitude}
\begin{figure}[!h]
\centerline{\includegraphics[width=12cm]{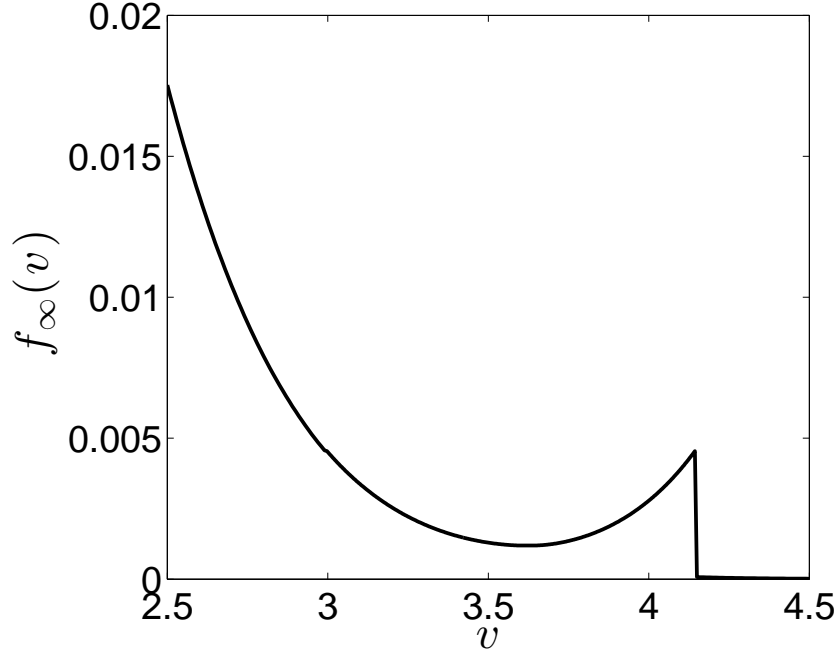}}
\caption{\label{finf} Electron distribution function corresponding to the situation when the EPW amplitude would have decreased to zero while its phase velocity would have converged towards $v_{\phi_\infty}\approx 3.62$. }
\end{figure}
If we assume that $dv_\phi/dv_{tr}$ remains bounded when the wave amplitude decreases down to small amplitudes, we prove in the Appendix that, 
\begin{equation}
\label{15}
\chi_a\sim \frac{4}{3\pi\sqrt{\Phi}}\left[2f_0(v_{\phi_0})-f_0(2v_\phi-v_{\phi_0})-Êf_0(3v_{\phi_0}-2v_\phi)Ê\right].
\end{equation}
Since $v_\phi$ does not converge back to $v_{\phi_0}$, this makes $\chi_a$ goes to infinity as $\Phi$ goes to zero, so that the equation $1+\chi_a=0$ can no longer be solved. Hence, the solutions to the adiabatic dispersion relation are such that $dv_\phi/dv_{tr}$ diverges when $v_{tr}Ê\rightarrow 0$. As discussed in the Appendix,  $v_\phi$ has to diverge logarithmically with $v_{tr}$ for $\chi_a$ to remain bounded, which is not physical for very small amplitudes. 

A logarithmic divergence of the wave frequency for small amplitudes was predicted in Ref.~\cite{dodin}, although the reason for such a divergence seems different from the one we are discussing here.

Numerically, we do find a divergence in $v_\phi$ at small amplitudes, which our numerical solver cannot really follow. Therefore, in Fig.õ\ref{fhys}, we choose not to show results corresponding to $\sqrt{\Phi}<0.05$ when the wave amplitude decreases. For such small amplitudes, we do not solve accurately the EPW dispersion relation, as may be seen in Fig.~\ref{knum}(b). \\

Let us now discuss in more detail the situation when the wave decays back to very small amplitudes.  If $v_\phi$ converged towards a finite value $v_{\phi_{\infty}}$ then, since  $\vert I \vert = v$ when $\Phi \rightarrow 0$,  using the results of Ref.~\cite{fadia} one could derive what would be the electron distribution function, $f_\infty(v)$. It is plotted in Fig.~\ref{finf} in the case when $v_{\phi_\infty}\approx 3.62$ (which corresponds to $\omega=1.195$).  It exhibits a sharp discontinuity, which is clearly not physical. However,  an abrupt transition in the distribution function has been observed numerically in Ref.~\cite{friou} and, since only the moments of the distribution function are physically relevant, the discontinuity in $f_\infty(v)$ is not an issue. Now, $df_\infty/dv>0$ in the velocity range $3.6 \alt v \alt 4.14$, so that most of the modes whose phase velocity, $v_{\phi_u}$, lies in that range grow unstable at the rate, 
\begin{equation}
\label{14}
\Gamma = \frac{\pi f'_\infty(v_{\phi_u})}{(k_u\lambda_D)^2\partial_\omega \chi_{lin_u}}.
\end{equation}
In Eq.~(\ref{14}),  $(k_u\lambda_D)$ is found by solving the linear dispersion relation with $f_\infty(v_\phi)$ as the unperturbed distribution function, and $\chi_{lin_u}$ is obtained by replacing $f_0$ by $f_\infty$ in Eq.~(\ref{A9}). The maximum growth rate is found for $v_{\phi_u} \approx 3.85$ and is $\Gamma/\omega_{pe}Ê\approx 3.3\times 10^{-2}$, so that $\Gamma/k\lambda_D\omega_{pe} \approx 0.1$. Now, adiabatic results are only accurate for EPW's whose growth rate is less than $0.1 (k\lambda_D)\omega_{pe}$, and only such slowly varying EPW's are considered in this article. This lets us conclude that, by the time the EPW has decreased to very small amplitudes, unstable modes have fully developed. Then, the electrostatic field in no longer monochromatic, and trying to derive its dispersion relation makes no sense.

\section{Frequency shift in a one-dimensional inhomogenous plasma}
\label{1D}

\subsection{Hypotheses}
\label{4.1}
There are several difficulties in deriving the EPW nonlinear dispersion relation in an inhomogeneous plasma. 

First, the advection of trapped electrons, at the phase velocity, changes the local electron density. Consequently, the ions can no longer be considered as a neutralizing background  and an electrostatic filed builds up in addition to that of the EPW. Here, we restrict to the situation when this electrostatic field is negligible, i.e., to the situation when the density of trapped electrons is very small. 

Second, if some electrons are detrapped, one has to account for their advection to correctly calculate the distribution function of the untrapped electrons. In order to avoid this difficulty, we restrict to the situation when the EPW keeps growing everywhere. Then, the density experienced by the EPW may be considered as a function of the wave amplitude, $n \equiv n(\Phi)$. Moreover, when deriving the action distribution function, we explicitly account for the conservation of the trapped electrons'  distribution function in the wave frame~\cite{benisti16,dodin12}. Hence, strictly speaking, the EPW nonlinear frequency is calculated in the frame moving at the local phase velocity. However, we checked that the values obtained for $\delta \omega$ did not change much if we simply assumed that the density of the trapped electrons was the same as that of the untrapped. Consequently, the values we find for $\delta \omega$ should be accurate in any reference frame. 

Third, in order to derive the nonlinear frequency shift one has to solve, self-consistently, for the change in the frequency and in the wavenumber. Just like in Section~\ref{hysteresis}, we neglect the change in $k$ due to the inhomogeneity in $\omega$. Then, $k$ is calculated a function of the density so that, in the linear regime, the EPW frequency remains constant. We checked that, for the parameters we used, this was a valid approximation. More precisely, for a given density profile, $n$, we calculate the wavenumbers, $k_{lin}$, such that $\omega_{lin}(k_{lin},n)=Const$. Then, we solve the nonlinear adiabatic dispersion relation with $k=k_{lin}$ to derive a first estimate of the nonlinear EPW frequency, $\omega_{NL}^{(1)} \{k_{lin},n,\PhiÊ\}$, where the braces indicate that $\omega_{NL}^{(1)}$ is a functional, and not a function, of $k_{lin}$, $\Phi$ and $n$. Using the profile thus found for $\omega_{NL}$, we solve the nonlinear dispersion relation for $k$ to derive $k_{NL}^{(1)} \{\omega_{NL}^{(1)}, \Phi, nÊ\}$, and we systematically find that $k_{NL}^{(1)}$ is close to $k_{lin}$. Then, we do not iterate the process and we assume that the nonlinear EPW frequency is $\omegaÊ\approx \omega_{NL}^{(1)}$. 

Fourth, as discussed above, when the wave keeps growing, the density may be considered as a function of the wave amplitude, $n\equiv n(\Phi)$. We restrict here to variations of $n$ with $\Phi$ such that the $l^{th}$-derivatives of $n$ with respect to $\sqrt{\Phi}$ increase less rapidly than $\Phi^{(1-l)/2}$ when $\Phi \rightarrow 0$. Then, following the results derived in the Appendix, $\chi_a$ converges towards a finite limit when $\Phi \rightarrow 0$.

\subsection{Derivation of the frequency shift and comparisons with local formulas}
\subsubsection{Several examples of the resolution of the nonlinear dispersion relation}
\begin{figure}[!h]
\centerline{\includegraphics[width=12cm]{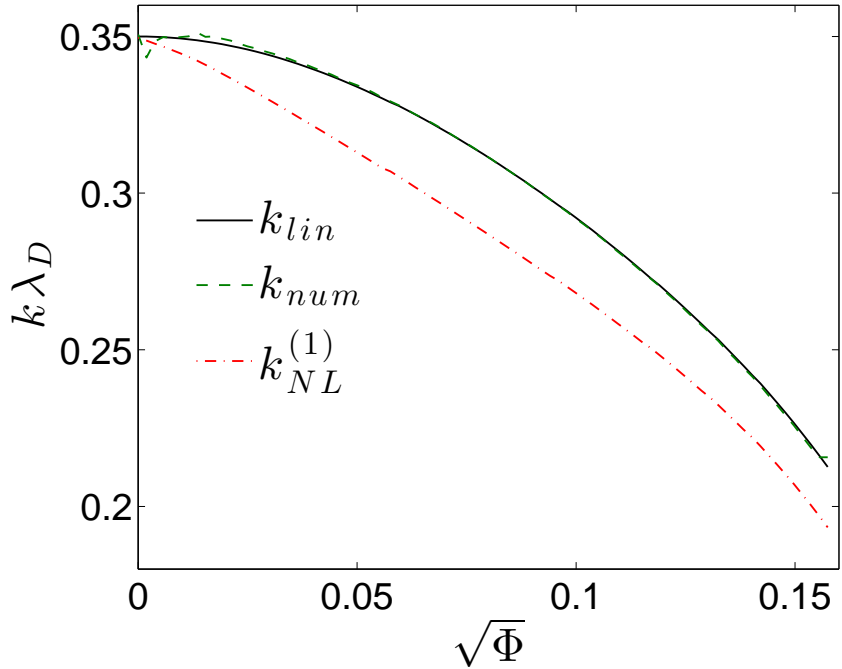}}
\caption{\label{f4} (Color online) The black solid line plots the values of $k_{lin}\lambda_D$ such that $\omega_{lin}(k_{lin},n)=1.21\omega_{pe}(0)$ when $n=n_0(1+11\Phi)$. The blue dashed line plots $k_{num}\lambda_D$ defined by Eq.~(\ref{10}). The red dashed-dotted line solves the nonlinear dispersion relation for $k$ when $n=n_0(1+11\Phi)$ and when the frequency varies with the amplitude as illustrated in Fig.~\ref{f5}.}
\end{figure}
In this Paragraph, we provide several results regarding the nonlinear variations of the EPW frequency, obtained by solving the dispersion relation~Eq.~(\ref{3}) for various density profiles, $n(\Phi)$. These are compared to the results obtained by using local formulas which are derived as follows. If one assumes that the wavenumber remains constant, $k \equiv k_0$, one may derive, like in Section~\ref{hysteresis}, the nonlinear variations of the EPW frequency while the wave is growing. Let us denote by $\omega_0(k_0,\Phi)$ the values thus obtained. Now, to a given density profile $n(\Phi)$ we associate a wavenumber profile $k_{lin}(\Phi)$ as explained in Paragraph~\ref{4.1}. Then, the so-called local estimate of the nonlinear EPW frequency is $\omega_{loc}(\Phi) \equiv \omega_0[k_{lin}(\Phi),\Phi]$. \\

\begin{figure}[!h]
\centerline{\includegraphics[width=12cm]{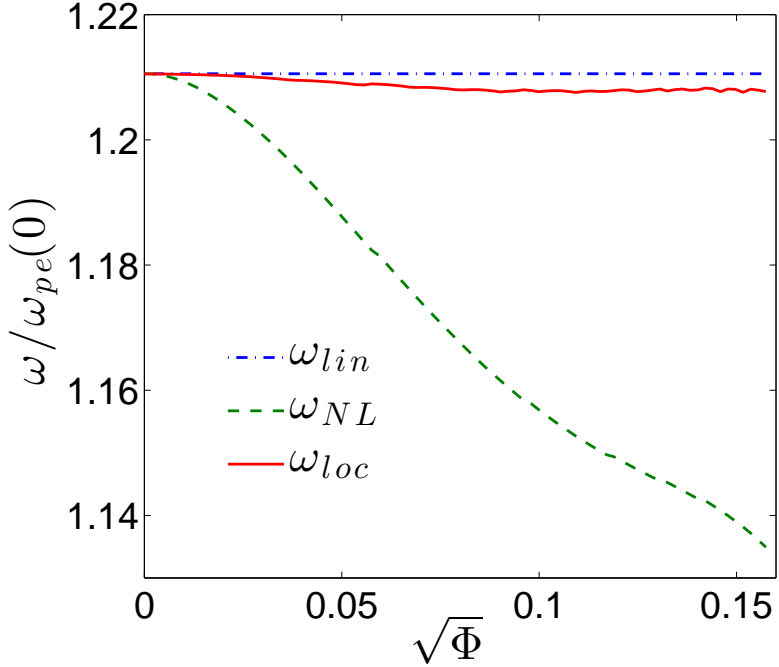}}
\caption{\label{f5} (Color online) EPW frequency, normalized to the initial plasma frequency, when $n=n_0(1+11\Phi)$ and when $k=k_{lin}$ such that $\omega_{lin}(k_{lin},n)=1.21\omega_{pe}(0)$. The blue dashed-dotted line plots the linear value of the frequency, the green dashed line plots the nonlinear EPW frequency solving Eq.~(\ref{3}), and the red solid line plots the nonlinear EPW frequency obtained by making use of local formulas.}
\end{figure}

Let us start with the situation when $n(\Phi)=n_0(1+11\Phi)$, $n_0$ being a constant. Then, the derivatives of $n$ with respect to $\sqrt{\Phi}$ remain bounded, as should be for $\chi_a$ to converge towards a finite limit when $\Phi \rightarrow 0$. The wavenumber profile is chosen so that, $\omega_{lin}(k_{lin},n)=1.21\omega_{pe}(0)$, where $\omega_{pe}(0)$ is the plasma frequency when $n=n(\Phi=0)$. Then, as shown in Fig.~\ref{f4}, when $\sqrt{\Phi}$ varies from 0 to 0.16, $k_{lin}\lambda_D$ varies from 0.35 to 0.21. Fig.~\ref{f5} plots the solution of the nonlinear dispersion relation Eq~(\ref{3}), $\omega(\Phi)  \equiv \omega_{NL}Ê\{k_{lin}(\Phi),n(\Phi),\Phi \}$, and compares it to the values obtained by using local formulas, $\omega_{loc}(\Phi) \equiv  \omega_0[k_{lin}(\Phi),\Phi]$. Although $\omega(\Phi)$ and $\omega_{loc}(\Phi)$ both decrease with the wave amplitude, the frequency shift found by solving the nonlinear dispersion relation is of much larger magnitude than that predicted by local formulas, which are clearly not accurate. 

Like in Section~\ref{hysteresis}, we make use of test particles simulations to check our resolution of Eq.~(\ref{3}). Hence, we numerically solve the equations of motion with $v_\phi=\omega_{NL}/k_{lin}$, $\omega_{NL}$ being given by green dashed line of Fig.~\ref{f5} and $k_{lin}$ by the black solid line of Fig.~\ref{f4}, and we compute $k_{num}$ as given by Eq.~(\ref{10}).  As may be seen in Fig.~\ref{f4}, we find $k_{num}=k_{lin}$which shows that our resolution of Eq.~(\ref{3}) is correct. In Eq.~(\ref{10}), we account for plasma inhomogeneity the following way. We calculate the density of the trapped electrons, $n_t$, by using the result $n_t/k=Const.$, derived in Refs.~\cite{benisti16,dodin12}. Then, the weight $p_i$ of trapped electrons is $p_i=n_te^{-v_{0_i}^2/2}/\sum_{j=1}^N p_j$. As regards the untrapped electrons, their weight is $p_i=ne^{-v_{0_i}^2/2}/\sum_{j=1}^N p_j$, $n$ being the local density.

In order to test the relevance of using $k=k_{lin}$ we solve Eq.~(\ref{3}) for $k$ with $\omega=\omega_{NL}$ given by the green dashed line in Fig.~\ref{f5}. As may be seen in Fig.~\ref{f4}, the values thus found for $k$ are indeed close to $k_{lin}$ (they differ from $k_{lin}$ by less than 10\%).  \\
\begin{figure}[!h]
\centerline{\includegraphics[width=12cm]{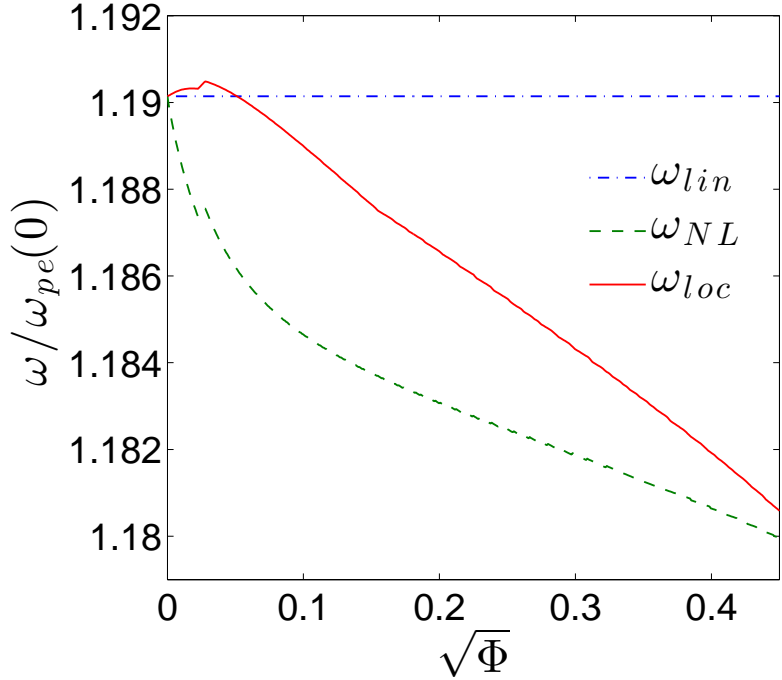}}
\caption{\label{f6} (Color online) EPW frequency, normalized to the initial plasma frequency, when $n=n_0(1+0.15\sqrt{\Phi}-0.2\Phi)$ and when $\omega_{lin}=1.19\omega_{pe}(0)$. The blue dashed-dotted line plots the linear value of the frequency, the green dashed line plots the nonlinear EPW frequency solving Eq.~(\ref{3}), and the red solid line plots the nonlinear EPW frequency obtained by making use of local formulas.}
\end{figure}

Let us now investigate another example, with smaller density variations, $n=n_0(1+0.15\sqrt{\Phi}-0.2\Phi)$, and $\omega_{lin}=1.19\omega_{pe}(0)$. When $\sqrt{\Phi}$ varies form 0 to 0.45, $k_{lin}$ first decreases from $k_{lin}\lambda_D=0.33$ down to $k_{lin}\lambda_DÊ\approx 0.3148$ when $\sqrt{\Phi} \approx 0.375$, and then slightly increases to $k_{lin}\lambda_DÊ\approx 0.3154$ when $\sqrt{\Phi}=0.45$. The wavenumber varies much less than in the previous example. Then, as may be seen in Fig.~\ref{f6}, there is a much better agreement between the values of $\omega$ derived form Eq.~(\ref{3}) and those inferred from local formulas. \\
\begin{figure}[!h]
\centerline{\includegraphics[width=12cm]{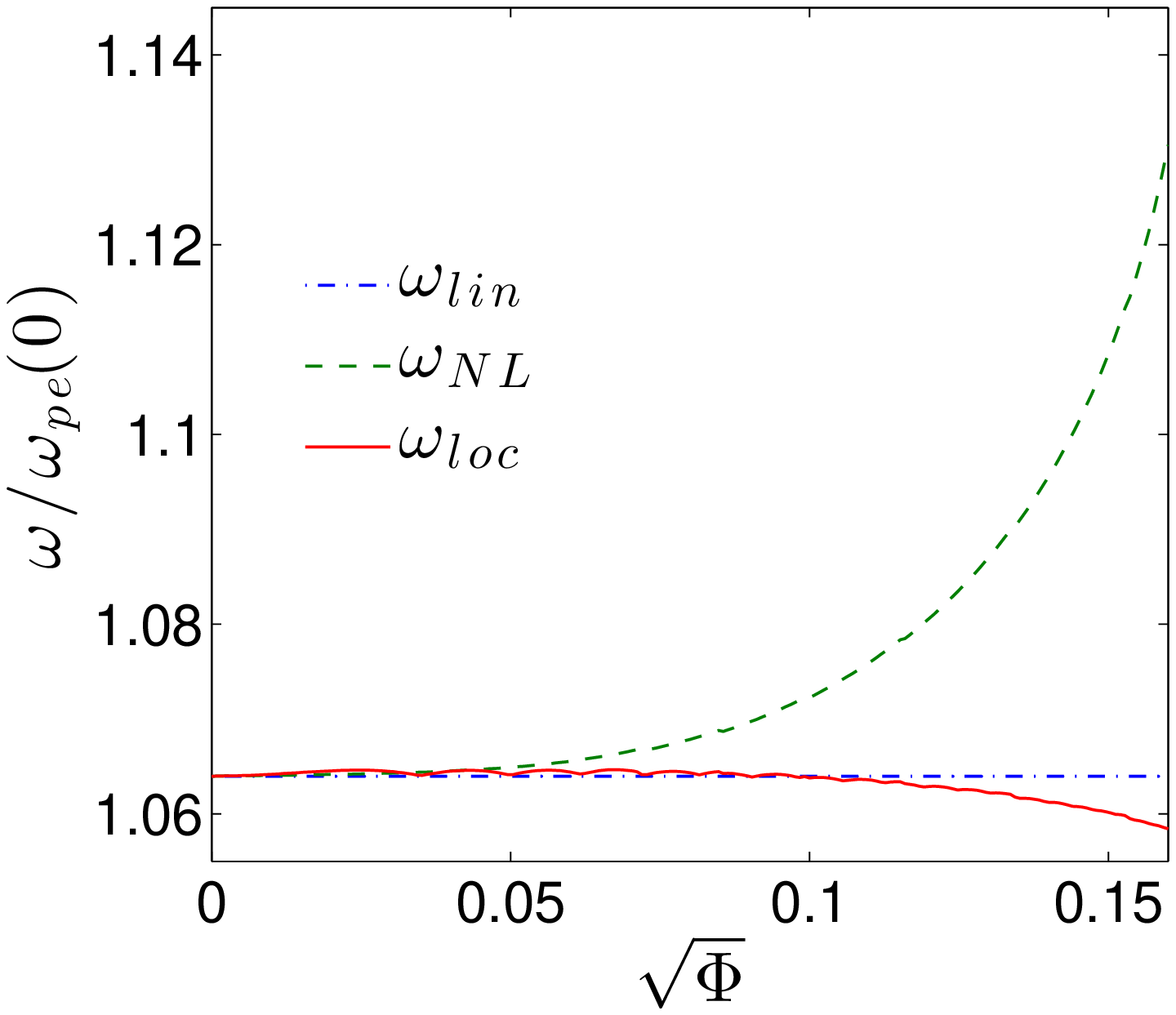}}
\caption{\label{f7} (Color online) EPW frequency, normalized to the initial plasma frequency, when  $n=n_0(1-11\Phi)$, and $\omega_{lin} \approx1.065\omega_{pe}(0)$. The blue dashed-dotted line plots the linear value of the frequency, the green dashed line plots the nonlinear EPW frequency solving Eq.~(\ref{3}), and the red solid line plots the nonlinear EPW frequency obtained by making use of local formulas.}
\end{figure}

Finally, let us investigate a situation when the density decreases, $n=n_0(1-11\Phi)$, and $\omega_{lin} \approx1.065\omega_{pe}(0)$, so that $k_{lin}$ increases from $k_{lin}\lambda_D= 0.2$ to $k_{lin}\lambda_D\approx 0.4$ when $\sqrt{\Phi}$ varies from $\sqrt{\Phi}=0$ to $\sqrt{\Phi}=0.16$. For this last exemple, the difference between $\omega_{NL}$ and $\omega_{loc}$ is striking. While one would predict a small negative frequency shift by making use of local formulas, solving Eq.~(\ref{3}) shows that the frequency shift is actually positive and of much larger amplitude. \\

\subsubsection{Discussion}

In an inhomogeneous plasma, the EPW phase velocity changes with time because of the nonlinear frequency shift, and because of the density variations. Then, it seems clear that local formulas should be valid when $v_\phi$ is more modified by nonlinearity than by plasma inhomogeneity. In order to determine what is  the main cause for the variations in $v_\phi$, one may use Eqs.~(\ref{13}) and (\ref{12}) to estimate the nonlinear shift, $\delta v_\phi$, in phase velocity. If $\delta v_\phi$ thus calculated is much larger than the variations in phase velocity entailed by inhomogeneity, local formulas should be valid. When $n=n_0(1+11\Phi)$ and $\omega_{lin}=1.21\omega_{pe_0}$ (which corresponds to the situation illustrated in Fig.~\ref{f5}), and when $n=n_0(1-11\Phi)$ and $\omega_{lin}=1.065\omega_{pe_0}$ (which corresponds to Fig.~\ref{f7}), $\delta v_\phi$ calculated when $\sqrt{\Phi}=0.16$ is much less than the variations in $v_\phi$ entailed by plasma inhomogeneity. Then, as expected, local formulas are not accurate, and even completely wrong in the situation illustrated in Fig.~\ref{f7}. When $n=n_0(1+0.15\sqrt{\Phi}-0.2\Phi)$ and $\omega_{lin}=1.19\omega_{pe}(0)$ (which corresponds to Fig.~\ref{f6}), $\delta v_\phi$ calculated when $\sqrt{\Phi}=0.45$ is close to the change in phase velocity due to inhomogeneity, and local formulas yield a fair approximation of the nonlinear variations in $\omega$. \\

\section{Frequency shift in a two-dimensional inhomogeneous plasma}
\label{2D}
\subsection{Hypotheses}
\label{5.1}
In this Section, we derive the nonlinear frequency sift of an EPW by accounting for two-dimensional (2-D) effects. We assume that the transverse profile (along the $y$-direction) of the wave electric field in a Gaussian, $\Phi(y) \equiv \Phi_0 e^{-2y^2/l_\bot^2}$, and that the transverse velocity distribution function is a Maxwellian, $f(v_y) \equiv e^{-v^2_y/2v_{\bot}^2}/\sqrt{2\pi}v_{\bot}$. Moreover, we assume that the wave grows exponentially in time, $\Phi(\tau) \equiv \Phi_0 e^{\tau/\tau_g}$. Then, like in Section~\ref{1D}, we can relate the density to the wave amplitude at $y=0$, $n \equiv n[\Phi(y=0)]$. The wave numbers are also calculated like in Section~\ref{1D}, $k\equiv k_{lin}$ such that $\omega_{lin}(k_{lin},n)=Const$.

Moreover, we restrict to  situations when the change in density is modest and when the number of trapped electrons is very small, so that their advection, or the advection of the electrons which are detrapped due to their transverse motion, induces a negligible change in the total charge density. 

Now, as discussed in Section~\ref{hysteresis}, solving Eq.~(\ref{3}) at the edge of $\mV$  provides values for $v_\phi$ that diverge logarithmically as $\Phi \rightarrow 0$. Consequently, we restrict our derivation of $\delta \omega$ to values of the EPW amplitude which are large enough for the solution of Eq.~(\ref{3}) to make sense. As discussed in Section~\ref{hysteresis}, due to detrapping the velocity distribution function assumes a positive slope, leading to the unstable growth of electrostatic modes at the edge of  the domain $\mV$. Hence, there is a range in $y$ where the electrostatic field in not monochromatic. We restrict our derivation of $\delta \omega$ to times so short that the unstable modes did not have the time to reach large amplitudes. Hence, in the space region where the electrostatic field is not monochromatic, its amplitude is too small to significantly modify the electron distribution function. This space region is simply neglected in our calculation. 

\subsection{Derivation of the nonlinear frequency shift. Comparisons with 1-D results}
\label{5.2}
\begin{figure}[!h]
\centerline{\includegraphics[width=12cm]{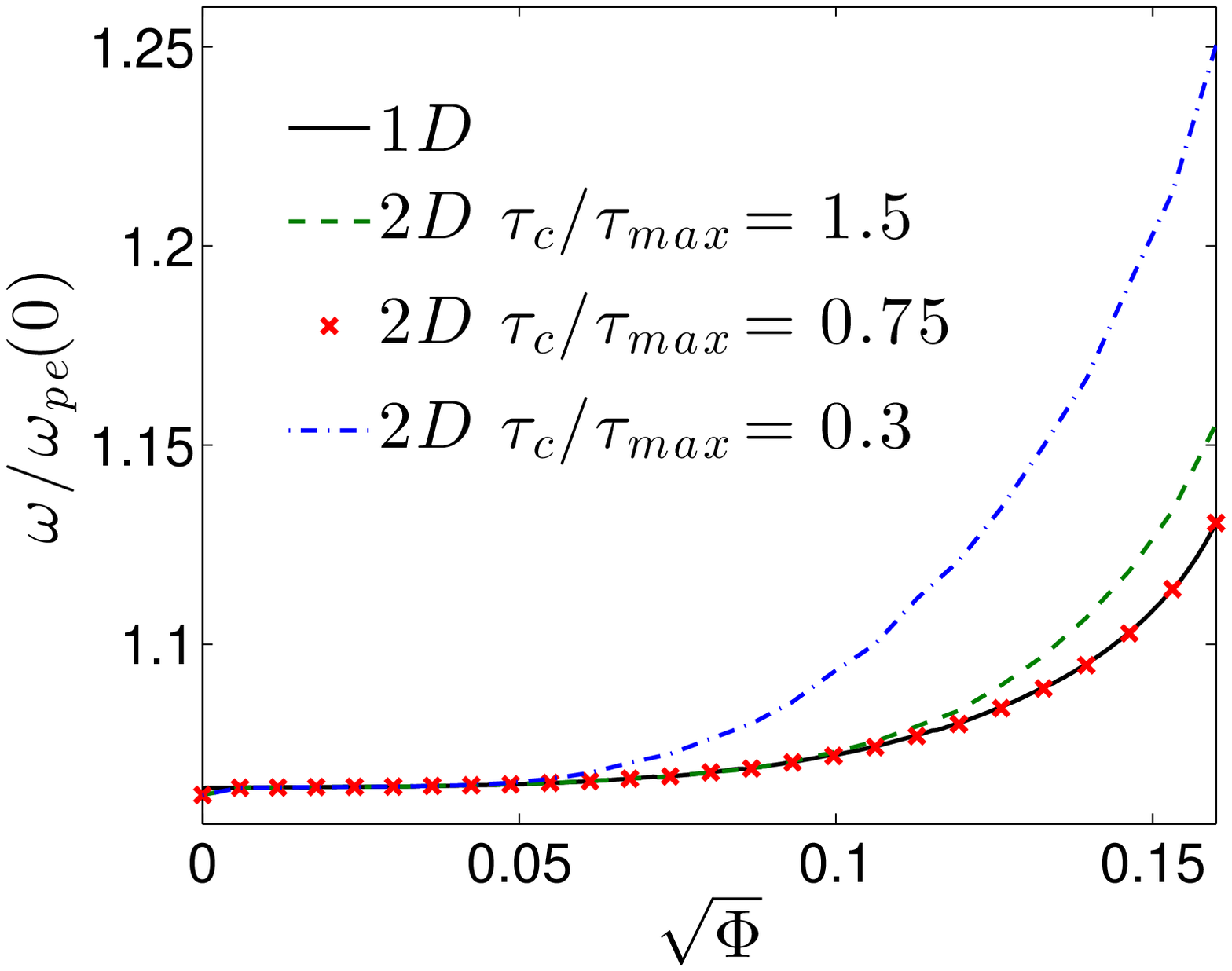}}
\caption{\label{f8} (Color online) EPW frequency calculated at $y=0$, normalized to the initial plasma frequency, when  $n=n_0(1-11\Phi)$, and $\omega_{lin} \approx1.065\omega_{pe}(0)$. The black solid line plots the 1-D results obtained in Section~\ref{1D}. The red crosses correspond to 2-D results when $\tau_c/\tau_{\max}=1.5$, the green dashed corresponds to $\tau_c/\tau_{\max}=0.75$ and the blued dashed-dotted line to $\tau_c/\tau_{\max}=0.3$.}
\end{figure}

In 2-D, and at any transverse position $y$, the dispersion relation reads $1+\chi_a^{2D}(y,\tau)=0$, where $\chi_a^{2D}$ is the two-dimensional value of the adiabatic electron susceptibility, and its expression straightforwardly follows from that derived in 1-D.  Indeed, let us first assume that all electrons have the same transverse initial velocity, $v_y$. Because $v_y$ is not affected by the wave, only the longitudinal electron motion is to be investigated. It follows from the equation, $dv/d\tau=-\Phi(y_0+v_y \tau,\tau) \sin(\varphi)$, where $y_0$ is the initial electron transverse position. Hence, if all electrons have the same initial position and transverse velocity, the adiabatic electron electron susceptibility, which we denote by $\chi_a^{1D}(y_0,v_y)$, is calculated exactly as in 1-D, provided that one uses  $\Phi(y_0+v_y \tau,\tau)$ for the wave amplitude. Then,
\begin{equation}
\label{n1}
\chi_a^{2D}(y,\tau)=\int \chi_a^{1D}(y-v_y\tau,v_y) f(v_y)dv_y. 
\end{equation}
Note that the motion of each class of electrons with transverse velocity, $v_y$, is adiabatic only when $l_\bot/v_y$ is large enough. For a Maxwellian distribution of transverse velocities, it is valid to use the adiabatic expression of $\chi_a^{1D}$ in Eq.~(\ref{n1}) provided that $\tau_c \equiv l_\bot/v_\bot$ is large enough (larger than about 10 when $v_\bot$ is normalized to the longitudinal thermal velocity and when $l_\bot$ is normalized to $k\lambda_D$), which we assume here. 

Let us not show specific examples of the nonlinear change in the EPW frequency, calculated at $y=0$, and let us compare these variations in $\omega$ with 1-D results. We only derive the frequency shift within a finite time interval, $0 \leq \tau \leq \tau_{\max}$, and the key parameter for our comparisons is the ratio $\tau_c/\tau_{\max}$. Since $\tau_c\equiv l_\bot/v_\bot$ is half of the typical time it takes for electrons to cross $\mV$, if $\tau_c /\tau_{\max}Ê\agt 1$ most electrons have not been detrapped due to their transverse motion, so that their response to the wave should be the nearly same as in 1-D, and 1-D values of $\delta \omega$ are expected to be accurate. As $\tau_c /\tau_{\max}$ decreases, more and more electrons are trapped and detrapped due to the transverse variation of $\Phi$, leading to the hysteresis in $\omega$ described in Section~\ref{hysteresis}. The main point of this Section is to discuss how the hysteresis modifies the nonlinear variations of the wave frequency in a two-dimensional geometry. \\

Fig.~\ref{f8} shows results corresponding to the same linear variations of the phase velocity as in Fig.~\ref{f7}. As expected, when $\tau_c/\tau_{\max}\agt1$,  $\omega$ varies with $\Phi$ in a  similar way as in 1- while, as $\tau_c/\tau_{\max}$ decreases, the discrepancy between 1-D and 2-D results keeps increasing. Moreover, we note that when $\tau_c/\tau_{\max}$ becomes smaller, the nonlinear values of $\omega$ become larger, which is a direct consequence of the hysteresis. Indeed, due to the continuous trapping and detrapping, the EPW frequency at the edge of the domain $\mV$ keeps on increasing with time. This means that, as time goes by, the electrons which enter the domain~$\mV$ experience a wave frequency whose value may significantly exceed the linear one. Hence, when deriving the EPW frequency shift in 2-D, we start with a small amplitude frequency which is larger then in 1-D and, quite logically, we find larger nonlinear frequencies, all the more as $\tau_c/\tau_{\max}$ is small. \\
\begin{figure}[!h]
\centerline{\includegraphics[width=12cm]{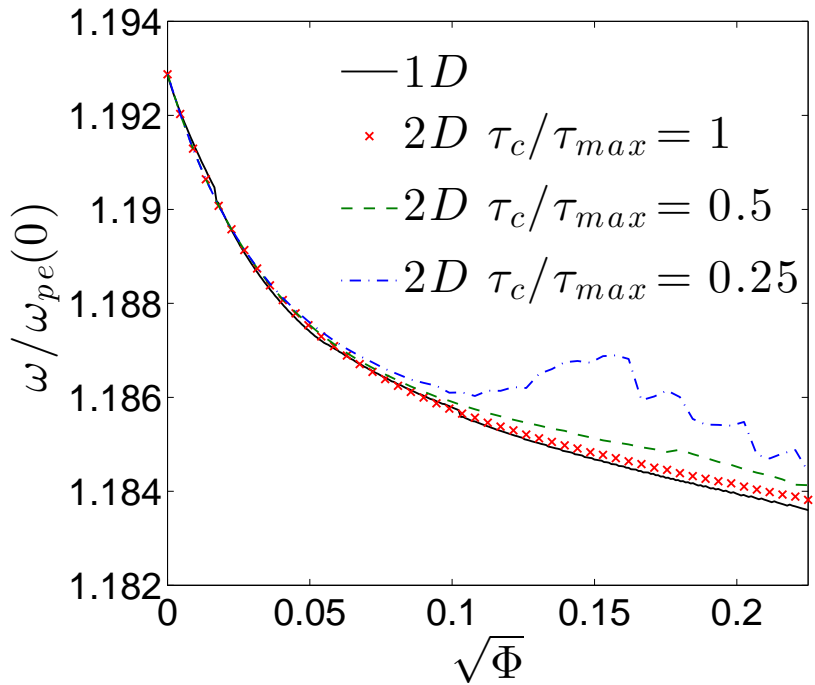}}
\caption{\label{f9} (Color online) EPW frequency calculated at $y=0$, normalized to the initial plasma frequency, when  $n=n_0(1+0.15\sqrt{\Phi}-0.2\Phi)$ and when $\omega_{lin}=1.193\omega_{pe}(0)$. The black solid line plots the 1-D results obtained in Section~\ref{1D}. The red crosses correspond to 2-D results when $\tau_c/\tau_{\max}=1$, the green dashed corresponds to $\tau_c/\tau_{\max}=0.5$ and the blued dashed-dotted line to $\tau_c/\tau_{\max}=0.25$.}
\end{figure}

The same trend may be observed in Fig.~\ref{f9}, which shows results corresponding to the same linear variations of the phase velocity as in Fig.~\ref{f6}. The agreement between 1-D and 2-D results remains good down to values of $\tau_c/\tau_{\max}$ as small as $\tau_c/\tau_{\max}=0.25$. Since we showed in Section~\ref{1D} that local formulas were fairly close to 1-D results, we conclude that this is one example where one may predict in advance the nonlinear variations of the frequency, regardless of how the wave grows. 

Note that, unlike in Fig.~\ref{f6} where $\sqrt{\Phi}<0.45$, in Fig.~\ref{f9} we restrict to wave amplitudes such that $\sqrt{\Phi}<0.225$. This is to make sure that, at the edge of the domain~$\mV$, the unstable modes do not have the time to grow significantly. Indeed, when $\sqrt{\Phi}=0.225$, the largest growth rate of these modes is close to $7\times10^{-3}\omega_{pe}$. 

\section{Conclusion}
\label{conclusion}
In this paper, we derived the nonlinear frequency variations of a sinusoidal EPW, in a two-dimensional inhomogeneous plasma. A particular emphasis was laid on the ability to correctly approximate these variations by formulas relating the frequency to the local wave amplitude. In order to discuss this point, we first addressed the situation when an EPW monotonously grew to a large amplitude, and then monotonously decayed to very small values. In this situation, we evidenced a hysteresis in the wave frequency which could not be neglected. This let us conclude that local formulas might only be valid for waves which essentially grew, in their reference frame. Moreover, we showed that solving the adiabatic nonlinear dispersion relation would lead to a logarithmic divergence in $\omega$ as the wave amplitude decayed back to small amplitudes. Physically, when $\Phi \rightarrow 0$, the velocity distribution function assumes a positive slope, leading to the unstable growth of electrostatic modes. Hence, there exists a minimum amplitude below which the electrostatic field may not be considered monochromatic, so that trying to derive its dispersion relation makes no sense. 

In a 1-D inhomogeneous plasma, we showed that the wave frequency depended on the whole history of the plasma density variations, relative to those of the EPW amplitude. Moreover, we showed that this was a direct consequence of the nonlocality, in $v_\phi$, of the action distribution function. This let us conclude that local formulas could only be accurate if the nonlinear variations in $v_\phi$ were larger than those induced by the plasma inhomogeneity. 

Finally, we addressed the derivation of $\delta \omega$ a 2-D inhomogeneous plasma, when electrons kept on being trapped and detrapped as they transversely crossed the domain, $\mV$, where the EPW electric field was significant. We showed that, because detrapping entailed a hysteresis in the wave frequency, 1-D results were only accurate before most electrons had the time to cross the domain $\mV$. \\

Hence, in summary, we conclude that local formulas for the EPW frequency shift may only be accurate for a time less than the typical time needed for electrons to cross the domain $\mV$ where the electric field is significant, provided that  the wave essentially grows in its reference frame, and that the nonlinear variations of the phase velocity are larger than those due to plasma inhomogeneity. 

As regards the application that motivated this work, as SRS-driven plasma wave usually keeps growing before saturation~\cite{friou}. Moreover, SRS preferentially grows in regions where the density is nearly uniform. Hence, the most stringent condition is on $\tau_c$. For a laser fusion, $\tau_c \sim 0.2$ps, while the time for SRS to saturate is of the order of several picoseconds~\cite{SRS3D}. This makes the use of local formulas for $\delta \omega$ questionable. 

\appendix
\section{Asymptotic values for the adiabatic susceptibility}
\label{A}
\setcounter{equation}{0}
\newcounter{app}
\setcounter{app}{1}

In this Appendix, we derive the adiabatic susceptibility in the limit of small wave amplitudes in two different physical situations. First, when the wave keeps growing from $\Phi \approx 0$, with an initial phase velocity, $v_\phi=v_{\phi_0}$. Second, when the wave amplitude keeps decreasing in a homogeneous plasma.

\subsection{The wave amplitude keeps growing from $\Phi \approx0$.}
\subsubsection{Contribution from the trapped electrons}
\label{aa}
From Eq.~(\ref{chit}), the contribution to $\chi_a$ from the trapped electrons is 
\begin{equation}
\label{A1}
\chi_t=\frac{1}{\Phi}Ê\int_0^{v_{tr}} f_\gamma(I)C_t(I)dI,
\end{equation}
where $C_t \equiv 2(k\lambda_D)^{-2}(-1+2E/K)$. Let us denote by $v_{tr}^*(I)$ [respectively $v_\phi^*(I)$] the value assumed by $v_{tr}$ (respectively $v_\phi$) when the orbit with action $I$ has been trapped. Then, in Eq.~(\ref{A1}) $I=v_{tr}^*(I)$.

If $\vert dv_\phi^*/dI\vert \leq 1$, using Eq.~(94) of Ref.~\cite{fadia},
\begin{equation}
\label{A2}
f_\gamma(I)=f_\alpha(I+v_\phi^*)(1+dv_\phi^*/dI)+f_\alpha(I-v_\phi^*)(1-dv_\phi^*/dI).
\end{equation}
In Eq.~(\ref{A2}), $f_\alpha$ and $f_\beta$ are normalized to unity. Hence, if we denote by $f_0$ the unperturbed velocity distribution function, by $n^*$ the electron density at the position when trapping occurred, and by $\langle n \rangle$ the averaged density at the current position, $f_\alpha(I)=n^*f_0(I)/\langle n \rangle$, and $f_\beta(I)=n^*f_0(-I)/\langle n \rangle$. Plugging these values for $f_\alpha$ and $f_\beta$ into Eq.~(\ref{A2}), and using the expression thus found for $f_\gamma$ in Eq.~(\ref{A1}) yields, when $\vert dv_\phi^*/dI\vert \leq 1$,
\begin{equation}
\label{A3}
\chi_t=\frac{1}{\Phi}Ê\int_0^{v_{tr}} \left[\frac{n^*}{\langle n \rangle} f_0(I+v_\phi^*)\left(1+\frac{dv_\phi^*}{dI} \right) +\frac{n^*}{\langle n \rangle} f_0(v_\phi^*-I)\left(1-\frac{dv_\phi^*}{dI} \right)\right]C_t(I)dI.
\end{equation}
If $dv_\phi^*/dI >1$, $f_\gamma=2f_\alpha$, so that,
\begin{equation}
\label{A4}
\chi_t=\frac{1}{\Phi}Ê\int_0^{v_{tr}} 2\frac{n^*}{\langle n \rangle} f_0(I+v_\phi^*)C_t(I)dI,
\end{equation}
while if $dv_\phi^*/dI <-1$, $f_\gamma=2f_\beta$ and,
\begin{equation}
\label{A5}
\chi_t=\frac{1}{\Phi}Ê\int_0^{v_{tr}} 2\frac{n^*}{\langle n \rangle} f_0(v_\phi^*-I)C_t(I)dI.
\end{equation}
Note that, since  the wave amplitude keeps growing, one may consider $n^*$ as a function of $v_{tr}^*$, $n^*\equiv n^*(v_{tr}^*)=n^*(I)$. 

Henceforth, for a reason that will become clear in a few lines, we only consider the situation when $\vert dv_\phi^*/dI \vert <1$. Making use of a Taylor expansion to second order in $I$, and using the identity proved in Ref.~\cite{benisti16}, $\Phi^{-1}\int_0^{v_{tr}} C_t(I)dI=8/3\pi\sqrt{\phi}(k\lambda_D)^2$, Eq.~(\ref{A3}) reads, 
\begin{eqnarray}
\nonumber
\chi_t &\approx &\frac{4f'_0(v_{\phi_0})}{\Phi}Ê\int_0^{v_{tr}} I \frac{dv_\phi^*}{dI}C_t(I)dI +\frac{2f_0(v_{\phi_0})}{\langle n \rangle \Phi}\int_0^{v_{tr}} (v_{tr}-I)\frac{dn^*}{dI}C_t(I)dI \\
\nonumber
&& + \frac{f^{''}(v_{\phi_0})}{\Phi} \int_0^{v_{tr}} I^2 \left[1+3\left(\frac{dv_\phi^*}{dI} \right)^2Ê\right]C_t(I)dI+\frac{4f'_0(v_{\phi_0})}{\langle n \rangle \Phi}\int_0^{v_{tr}}I(v_{tr}-I)\frac{dn^*}{dI}\frac{dv_\phi^*}{dI}C_t(I)dI\\
\label{A6}
&&+\frac{f_0(v_{\phi_0})}{\langle n \rangle\Phi}\int_0^{v_{tr}} (v_{tr}-I)^2\frac{d^2n^*}{dI^2}C_t(I)dI+\frac{16 f_0(v_{\phi_0})}{3\pi\sqrt{\Phi}(k\lambda_D)^2},
\end{eqnarray}
where $f'_0 \equiv df_0/dv$ and $f^{''}_0 \equiv d^2f_0/dv^2$.
\subsubsection{Contribution from the untrapped electrons}

From Eq.~(\ref{chiu}), when $\vert dv_\phi^*/dI \vert <1$ and when the fraction of trapped electrons is negligible, the contribution to $\chi_a$ from the untrapped electrons is,
\begin{equation}
\label{A7}
\chi_u=\frac{1}{\Phi}Ê\int_{v_{tr}}^{+\infty}\left[f_0(v_\phi+I)+f_0(v_\phi-I)-2f_0(v_\phi)Ê\right]C_u(I)dI-\frac{16 f_0(v_{\phi})}{3\pi\sqrt{\Phi}(k\lambda_D)^2}, 
\end{equation}
where we have denoted $C_u\equiv 2(k\lambda_D)^{-2}[1+(2/m)(E/K-1)]$, and where we have used the identity proved in Ref.~\cite{benisti16}, $\Phi^{-1}\int_0^{v_{tr}} C_u(I)dI=-8/3\pi\sqrt{\phi}(k\lambda_D)^2$. Now, as shown in Ref.~\cite{benisti16}, in the limit when $I/v_{tr}Ê\rightarrow \infty$, 
\begin{equation}
\label{A8}
C_u(I) \approx \frac{-\Phi}{(k\lambda_D)^2I^2}.
\end{equation}
Taking advantage of the latter identity, we decompose the integral in Eq.~(\ref{A8}) into two contributions, one when $I$ varies from $v_{tr}$ to $A v_{tr}$ and the other when  $I$ varies from $A v_{tr}$ to $+\infty$. Here, $A$ is chosen large enough for Eq.~(\ref{A8}) to be valid when $I>Av_{tr}$, and small enough for a Taylor expansion of $f_0(v_\phi \pm I)$ to be valid when $I<Av_{tr}$. Then, making use of the change of variables, $I=v_{tr}u$, Eq.~(\ref{A7}) reads,
\begin{eqnarray}
\chi_u \approx \frac{f^{''}_0(v_\phi)}{(k\lambda_D)^2} \sqrt{\Phi} \left[ \frac{64(k\lambda_D)^2}{\pi^3}\int_1^A u^2C_udu+\frac{4A}{\pi}ÊÊ\right]+\chi_{lin}-\frac{16 f_0(v_{\phi})}{3\pi\sqrt{\Phi}(k\lambda_D)^2}, 
\end{eqnarray}
where $\chi_{lin}$ is the linear electron susceptibility, 
\begin{equation}
\label{A9}
\chi_{lin}\equiv P.P.\left(-\frac{1}{(k\lambda_D)^2}\int \frac{f_0(v)-f_0(v_\phi)}{(v-v_\phi)^2}dv\right).
\end{equation}
The integral, $\int_1^A u^2C_udu$ is evaluated numerically, and we find that the right-hand side of Eq.~(\ref{A9}) becomes essentially independent of $A$ when $A>3$, and is,
\begin{equation}
\label{A10}
\chi_u \approx -\frac{16 f_0(v_{\phi})}{3\pi\sqrt{\Phi}(k\lambda_D)^2}+\chi_{lin}+\frac{1.125\sqrt{\Phi}f^{''}_0(v_\phi)}{(k\lambda_D)^2}.
\end{equation}

\subsubsection{Asymptotic value of the adiabatic susceptibility, and of the nonlinear frequency shift}

From Eqs.~(\ref{A6}) and (\ref{A10}), $\chi_a\equiv \chi_t+\chi_u$ is,
\begin{eqnarray}
\nonumber
\chi_a & \approx &\frac{16 [f_0(v_{\phi_0})-f_0(v_{\phi})]}{3\pi\sqrt{\Phi}(k\lambda_D)^2}+ \chi_{lin}+\frac{1.125\sqrt{\Phi}f^{''}_0(v_\phi)}{(k\lambda_D)^2} \\
\nonumber
&&+\frac{4f'_0(v_{\phi_0})}{\Phi}Ê\int_0^{v_{tr}} I \frac{dv_\phi^*}{dI}C_t(I)dI +\frac{2f_0(v_{\phi_0})}{\langle n \rangle \Phi}\int_0^{v_{tr}} (v_{tr}-I)\frac{dn^*}{dI}C_t(I)dI \\
\nonumber
&&+ \frac{f^{''}(v_{\phi_0})}{\Phi} \int_0^{v_{tr}} I^2 \left[1+3\left(\frac{dv_\phi^*}{dI} \right)^2Ê\right]C_t(I)dI+\frac{4f'_0(v_{\phi_0})}{\langle n \rangle \Phi}\int_0^{v_{tr}}I(v_{tr}-I)\frac{dn^*}{dI}\frac{dv_\phi^*}{dI}C_t(I)dI \\
\label{A11}
&& +\frac{f_0(v_{\phi_0})}{\langle n \rangle\Phi}\int_0^{v_{tr}} (v_{tr}-I)^2\frac{d^2n^*}{dI^2}C_t(I)dI.
\end{eqnarray}
Now, it is clear that $\chi_a$ converges towards a finite limit when $v_{tr}\rightarrow 0$ provided that $dn^*/dI$ and $Id^2n^*/dI^2$ (and more generally $I^{l-1}d^ln^*/dI^l$) remain bounded when $I\rightarrow 0$ (while $dv_\phi^*/dI$ is necessarily bounded since we assumed $\vert dv_\phi^*/dI \vert<1$). When this condition is fulfilled, the lowest order nonlinear correction to $\chi_a$ is proportional to $\sqrt{\Phi}$. For a homogeneous plasma this entails that $v_\phi^*(I)-v_{\phi_0} \sim \eta_v I$ when $I\rightarrow 0$, where $\eta_v$ is a constant.  Moreover, in a few lines, we will show that $\eta_v \ll 1$, which vindicates our choice to focus on the situation when $\vert dv_\phi^*/dI \vert <1$. 

Let us now specialize to a homogeneous plasma.  By making use of the change of variables $I=v_{tr}u$ and of Taylor expansions to express $f_0(v_{\phi})$ in terms of $f_0(v_{\phi_0})$, one finds at lowest order in $\Phi$,  
\begin{eqnarray}
\nonumber
\chi_a & \approx & \chi_{lin}+\frac{64}{\pi^2}\eta_vf'_0(v_{\phi_0})Ê\left[\int_0^{1} uC_tdu -\frac{1}{3(k\lambda_D)^2}Ê\right]  \\
\label{corrige}
&&+\frac{\sqrt{\Phi}f^{''}_0(v_{\phi_0})}{(k\lambda_D)^2}\left[1.125+\frac{64}{\pi^3} (1+3\eta_v^2Ê)(k\lambda_D)^2 \int_0^{1} u^2 C_tdu-\frac{128 \eta_v^2}{3\pi^3} \right]\
\end{eqnarray}
Evaluating the integrals numerically yields the following approximate value for $\chi_a$, 
\begin{equation}
\label{A13}
\chi_a \equiv \chi_0 +\sqrt{\Phi}Ê\delta \chi,
\end{equation}
with
\begin{eqnarray}
\label{A14}Ê\
\chi_0& \approx &\chi_{lin}-1.5 \eta_v f'_0(v_{\phi})/(k\lambda_D)^2, \\
\delta \chi & \approx &  (1.09+4.84\eta_v^2)f^{''}_0(v_{\phi})/(k\lambda_D)^{2}.
\end{eqnarray}
Note that, $\lim_{\Phi \rightarrow 0} \chi_a =\chi_0 \neq \chi_{lin}$. Consequently, $v_{\phi_0}Ê\neq v_{\phi_{lin}}$. This is one defect of adiabatic formulas, they do not converge towards the linear limit when $\Phi \rightarrow 0$. However, the difference between $\chi_0$ and $\chi_{lin}$ is very small so that, in practice, $v_{\phi_0}Ê\approx v_{\phi_{lin}}$. 

Let us now provide an approximate solution to the dispersion relation, in the limit $\Phi \rightarrow 0$. To do so, we use the expansion, $v_{\phi}Ê\approx v_{\phi_0}+\eta_v v_{tr}$. Then, the dispersion relation $1+\chi_a=0$ reads,
\begin{eqnarray}
\nonumber
\eta_v &\approx& \frac{-\pi}{4}\frac{ \delta \chi}{\partial_{v_\phi}Ê\chi_0} \\
\label{A15}
& \approx & -\frac{(0.86 +3.80 \eta_v^2)f^{''}_0(v_{\phi_0})}{(k\lambda_D)^{2}\partial_{v_{\phi}}Ê\chi_0}.
\end{eqnarray}
When $f_0$ is a Maxwellian, the expression for $\chi_{lin}$ is known~\cite{gould}
\begin{equation}
\label{A16}
(k\lambda_D)^2\chi_{lin}=1-v_\phi e^{-v_\phi^2/2}Ê\int_0^{v_\phi}e^{u^2/2}du,
\end{equation}
which, using $1+\chi_{lin}(v_{\phi_{lin}})=0$, yields,
\begin{equation}
\label{A17}
(k\lambda_D)^2\partial_{v_\phi}\chi_{lin}=\frac{(\omega_{lin}/\omega_{pe})^2-1-(k\lambda_D)^2}{v_{\phi_{lin}}}.
\end{equation}
Therefore, Eq.~(\ref{A16}) reads,
\begin{equation}
\label{A18}
\eta_v \approx  -\frac{v_{\phi_{lin}}Ê(0.86 +3.80 \eta_v^2)f^{''}_0(v_{\phi_0}) }{(\omega_{lin}/\omega_{pe})^2-1-(k\lambda_D)^2-1.5 \eta_vv_{\phi_{lin}} f^{''}_0(v_{\phi})}.
\end{equation}

\subsection{The wave amplitude decreases towards $\Phi \approx 0$ in a homogeneous plasma}
When the wave amplitude keeps on decreasing, the contribution to $\chi_a$ from the trapped electrons assumes the same expression as that derived when the wave grows. Hence, when the plasma is homogeneous,
\begin{equation}
\label{A17}
\chi_t \equiv \delta \chi_t+\frac{16 f_0(v_{\phi_0})}{3\pi \sqrt{\Phi}(k\lambda_D)^2},
\end{equation}
with 
\begin{equation}
\label{A18}
\delta \chi_t=\frac{1}{\Phi}Ê\int_0^{v_{tr}} \left[ f_0(I+v_\phi^*)\left(1+\frac{dv_\phi^*}{dI} \right) +f_0(v_\phi^*-I)\left(1-\frac{dv_\phi^*}{dI} \right)-2f_0(v_{\phi_0})\right]C_t(I)dI.
\end{equation}
From the results obtained in Paragraph \ref{aa}, we know that $\delta \chi_t$ converges towards a finite limit when $\Phi \rightarrow 0$.

As regards the untrapped electrons, using Eqs.~(\ref{chiu}), (\ref{7}) and (\ref{7b}), one finds,
\begin{equation}
\label{A19}
\chi_u \equiv \delta\chi_{u_1}+\delta\chi_{u_2}+\delta\chi_{u_3}+\delta\chi_{u_4}-\frac{4}{3\pi \sqrt{\Phi}(k\lambda_D)^2}\left[2f_0(v_{\phi_0})+f_0(2v_\phi-v_{\phi_0})+Êf_0(3v_{\phi_0}-2v_\phi)Ê\right],
\end{equation}
where
\begin{eqnarray}
\label{A20}
\delta \chi_{u_1} &=& \frac{1}{2\Phi}Ê\int_{v_{tr}}^{+\infty} \left[f_0(I+v_\phi+v_\phi^*-v_\phi^{'*})-f_0(v_{\phi_0})Ê\right]C_udI, \\
\label{A21}
\delta \chi_{u_2} &=& \frac{1}{2\Phi}Ê\int_{v_{tr}}^{+\infty} \left[f_0(v_\phi^{'*}+v_\phi^*-v_\phi-I)-f_0(v_{\phi_0})Ê\right]C_udI, \\
\label{A22}
\delta \chi_{u_3} &=& \frac{1}{2\Phi}Ê\int_{v_{tr}}^{+\infty} \left[f_0(I+v_\phi+v_\phi^{'*}-v_\phi^*)-f_0(2v_\phi-v_{\phi_0})Ê\right]C_udI, \\
\label{A23}
\delta \chi_{u_4} &=& \frac{1}{2\Phi}Ê\int_{v_{tr}}^{+\infty} \left[f_0(3v_\phi^*-v_\phi^{'*}-v_\phi-I)-f_0(3v_{\phi_0}-2v_\phi)Ê\right]C_udI,
\end{eqnarray}
where we recall that $v_\phi^*$ is the wave phase velocity when the orbit has been trapped, and $v_\phi^{'*}$ is the EPW phase velocity when the orbit has been detrapped. From the results obtained when the wave was growing, we know that $dv_\phi^*/dI$ remains bounded when $I \rightarrow 0$. If this is also true for $dv_\phi^{'*}/dI$, each $\delta \chi_{u_i}$ ($1\leq i \leq4$) remains bounded when $\Phi \rightarrow 0$. Then, for small wave amplitudes,
\begin{equation}
\label{A24}
\delta \chi_a \sim \frac{4}{3\pi \sqrt{\Phi}(k\lambda_D)^2}\left[2f_0(v_{\phi_0})-f_0(2v_\phi-v_{\phi_0})-Êf_0(3v_{\phi_0}-2v_\phi)Ê\right].
\end{equation}
Now, since $v_\phi$ does not converge back to $v_{\phi_0}$, the right-hand side of Eq.~(\ref{A24}) diverges as $1/\sqrt{\Phi}$ when $\Phi \rightarrow 0$. The only way to cancel out this divergence with the $\delta \chi_{u_i}$'s is to let $dv_\phi^{'*}/dI$ diverge as $1/I$, which means that $v_\phi$ should diverge logarithmically when $\Phi$ decreases to zero.

\begin{acknowledgments}
The author would like to thank I.Y. Dodin for useful discussions and the unknown referee for pointing out Refs.~\cite{kis1,kis2}. 
 \end{acknowledgments}


\begin{thebibliography}{10}
\bibitem{explication}
There was, a priori, no reason to stop at order 12, except the memory limitations we had to face when running our routine for symbolic computation. However, going to order 12 was enough for the purpose of this paper, and, therefore, we did not try to go beyond this order. As regards the computation of $\langle \sin(x) \rangle$, only odd orders give a nonzero contribution so that its value at order 11 is the same as that obtained at order 12.
\bibitem{explication2}
When plotting the perturbative estimate of $\langle \sin(x) \rangle$ as a function of $\Phi$, we actually only find a local maximum when $I_0\alt 1.2$, which is a defect of the perturbative expansion. When $I_0\agt1.2$, the first maximum of $\langle \sin(x) \rangle$, and the amplitude when it is reached, are estimated by the point where the slope of $\langle \sin(x) \rangle(\Phi)$, as calculated perturbatively, is minimum. This estimate remains accurate as long as $I_0 \alt 1.6$, and, for these values of $I_0$, it does provide a good estimate of the jump in action, as shown in Section~\ref{III}.
\end{thebibliography}
\end{document}